# Recent Trend of Nanotechnology Applications to Improve Bio-accessibility of Lycopene by Nanocarrier: A Review

Anwar Ul Alam, M.[1]*, Khatun, M.[2], Arif Ul Alam, M.[3,4]
[1]Postdoctoral Associate, Dept. of Food Science, Cornell University, Ithaca, NY, USA. Email: ma989@cornell.edu
[2]Graduate Student, Dept. of Political Science and Public Administration, University of Alabama, Birmingham, USA, Email: mhamuda4@uab.edu
[3]Assistant Professor, Dept. Computer Science, University of Massachusetts Lowell and
[4]Assistant Professor (Adjunct), Dept. of Medicine University of Massachusetts Chan Medical School, Email: mohammad_alam@uml.edu

* = Corresponding author

**Abstract**

Lycopene, rich in red, yellow, or orange-colored fruits and vegetables, is the most potent antioxidant among the other carotenoids available in human blood plasma. It is evident that regular lycopene intake can prevent chronic diseases like cardiovascular diseases, type-2 diabetes, hypertension, kidney diseases and cancer. However, thermal processing, light, oxygen, and enzymes in gastrointestinal tract (GIT) compromise the bioaccessibility and bioavailability of lycopene ingested through diet. Nanoencapsulation provides a potential platform to prevent lycopene from light, air oxygen, thermal processing and enzymatic activity of the human digestive system. Physicochemical properties evidenced to be the potential indicator for determining the bioaccessibility of encapsulated bioactive compounds like lycopene. By manipulating the size or hydrodynamic diameter, zeta potential value or stability, polydispersity index or homogeneity and functional activity or retention of antioxidant properties observed to be the most prominent physicochemical properties to evaluate beneficial effect of implementation of nanotechnology on bioaccessibility study. Moreover, the molecular mechanism of the bioavailability of nanoparticles is not yet to be understood due to lack of comprehensive design to identify nanoparticles' behaviors if ingested through oral route as functional food ingredients. This review paper aims to study and leverage existing techniques about how nanotechnology can be used and verified to identify the bioaccessibility of lycopene before using it as a functional food ingredient for therapeutic treatments.

**Keywords:** Nanoencapsulation, lycopene, nanoparticles, bioaccessibility, control release kinetics.

**Contribution:**

**Mohammad Anwar Ul Alam:** Conceptualization, Data curation, Investigation, Methodology, Project administration, Supervision, Visualization, Writing original draft.
**Mhamuda Khatun:** Validation, Review and editing
**Mohammad Arif Ul Alam:** Review and editing

**Funding**

This work is partially supported by NSF's Smart & Connected Community award #2230180

## 1. Introduction

Lycopene is a 40 carbon long hydrophobic non-provitamin A carotenoid compounds contributing red or yellow color to the fruits and vegetables (1). Its chemical structure contains two nonconjugated and eleven conjugated double bonds which exhibits a distinct powerful antioxidant activity capable of preventing chronic diseases like cancer and cardiovascular diseases if ingested through diet (2, 3). But its antioxidant properties compromised by light, heat, and temperature due to oxidation and cis trans isomerization reaction (Figure 1) (4). Bio-accessibility, measured through antioxidant property analysis, of lycopene also enhances by dietary fat intake as it helps the lycopene to bind with chylomicrons which is a lipoprotein carries lycopene from intestinal enterocytes to lymphatic or portal circulation (5). Additionally, β-carotene oxygenase 1 (BOC1) and β-carotene oxygenase 2 (BOC2) present in intestinal enterocyte causes partial degradation of lycopene before absorption and thus influences its bio-accessibility and bioavailability (6). Therefore, encapsulating lycopene in biodegradable polymer attracts considerable interest, as it protects lycopene from degradation and breakdown into biological metabolites without producing any toxicity into the human body metabolism (7, 8, 9). Encapsulation at nano-level (size range from 1 to 1000 nm) also provides extra protection of the bioactive compounds like lycopene against the adverse processing and enzymatic degradation effects which further enhance bio-accessibility, bioavailability and controlled release to the targeted cells (10).

Nanotechnology is the study of nanoparticles (NPs) usage at the nanoscale level with a size range from 1 to 1000 nanometers (nm) (11, 12, 13, 14). Nanoparticles are categorized into different types based on their morphology, size, chemical and physical properties. Polymeric NPs exhibit an advanced platform for a wide array of biomedical interventions due to its non-toxigenic effect in body metabolism (15, 16, 17). On the other hand, engineered NPs can be used in multidimensional biomedical applications such as bio-imaging, bio-sensing, therapeutic delivery, and tissue engineering (16, 18, 19). Because of their size, bioactive compounds encapsulated in polymeric NPs can be diffused through the cellular barrier of tiny capillaries to get the maximum bioaccessibility or delivery efficiency for the targeted organs. The water insoluble biomaterials (lycopene) specially been extensively carried through biodegradable polymeric NPs in the biological system to improve their absorption capacity through the intestinal enterocytes (12, 13, 8, 15, 16, 18, 20, 21).

Various techniques have been adopted to synthesize encapsulated NPs based on their applications (22, 23, 21, 24). Optimization of composition and formulation methods help to identify the desired size of the nanospheres suitable for biomedical application (25, 26, 27, 28). Another property influencing the formulation process is interaction of bioactive compounds and polymers in a polymeric NPs (29, 30, 8, 31). The emulsification solvent-evaporation technique was identified as the most widespread method to formulate polymeric NPs (13, 32, 33). The biodegradable polymer is used frequently as a carrier for bioactive compounds to improve its bioavailability, encapsulation efficiency, controlled release kinetics, and targeted delivery to the organ without producing any toxicity into the body metabolism (34, 29, 35, 36). Physicochemical properties like zeta potential or stability, particle size, and size distribution are some distinct characteristics that determine toxicity effects, better biological functionality, and targeting ability of NPs in-vivo (37, 38, 39, 40, 19, 41, 42, 13). Many studies have investigated that nano-sized particles have several benefits over micro-particles-based drug delivery systems (43, 44, 45) due to their higher intracellular uptake compared to microparticles (22, 46, 47, 36).

Introduction of encapsulated bioactive compound's benefits to human health is a challenging effort. The food-based approach is always better to address a nutritional problem than a therapeutic approach due to its high acceptability in the consumer markets. To address nutritional problems, fruit juice has been used as a traditional fortification media for different well-accepted vitamins (like water soluble vitamin C and fat-soluble vitamin A and D) and minerals (i.e., iron, Ca) for many years due to its higher consumption profile among the consumers (48, 49). However, fortification of encapsulated lycopene to fruit juices to transfer its benefits to the diversified consumer will compromise as it might affect the physicochemical properties of the fortified juices and thus decrease its acceptability. Additionally, different food processing methods influence the bio-accessibility of encapsulated bioactive compounds fortified in fruit juices (50, 51). Researcher evidenced that, the thermal processing such as conventional pasteurization, microwave pasteurization, pulse electric field, and pulse UV light impact the quality of fruit juices, including the total soluble solids (TSS), pH, color, particle size, antioxidant content, total phenolic compounds, and titratable acidity of the products (52, 53). Subsequently, processing treatments might also affect the physicochemical properties of the encapsulated NPs. Researchers evaluated that if processing can change the hydrodynamic diameter, zeta potential, polydispersity index, rheological properties, thermal properties, and particle size distribution, conferred as physicochemical properties of nanoemulsion, have a definite effect on the control release activity and bioaccessibility of encapsulated NPs (54, 55). In Spite of this, enzymes catalyze polymers in the human GIT system and thus affect the particle size, antioxidant activity, and biodegradation rate of the biodegradable polymer used in encapsulating the bioactive compounds and thus affect the control release of bioactive compounds (56). Therefore, this study aims to show how nanotechnology can be applied for improving bioaccessibility and release kinetics of lycopene into the human GIT.

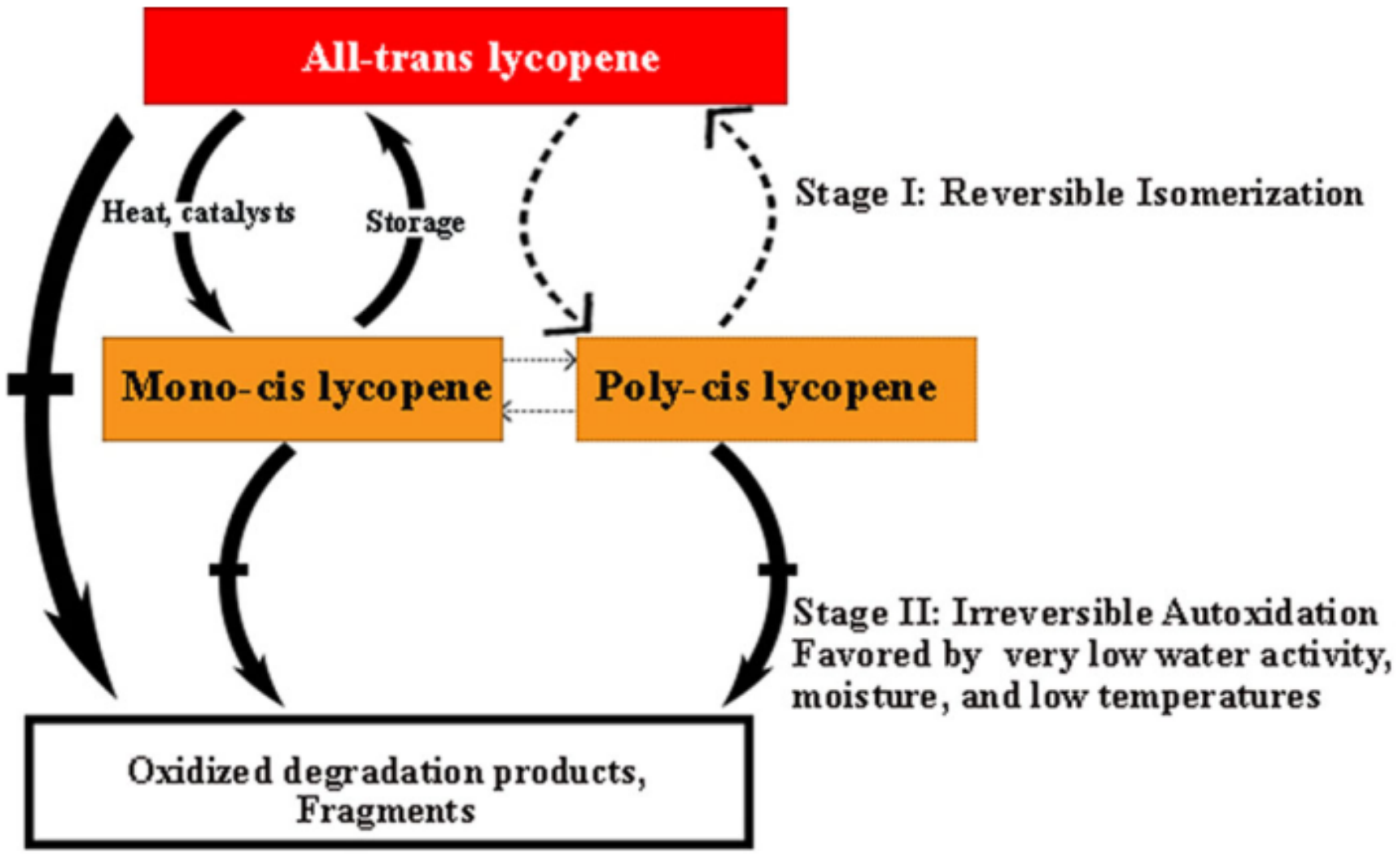

Figure 1. Degradation of lycopene due to heat, oxygen and light. Adopted from Tanambell et al. (57)

**2. Lycopene and chronic diseases:**

Carotenoids are lipophilic compounds with a structural backbone of poly-isoprenoid. Most carotenoids contain conjugated double bonds, which susceptible to cis-trans isomerization and oxidative degradation (Figure 1). Researchers identified six carotenoids (β-cryptoxanthin, lutein, zeaxanthin, β-carotene, α-carotene and lycopene) in human body fluid (i.e. human plasma and tissue). An extensive study has been done with β-carotene but not with lycopene. Very recently, lycopene has drawn significant attention due to its correlation with reducing the risk of chronic diseases, including cardiovascular diseases, type-2 diabetes, kidney diseases and cancers. Therefore, currently biological, and physicochemical properties of lycopene are extensively observed in different ongoing research projects. Molecular mass and chemical formula are same for lycopene and β-carotene with exception of lacking the β-ionone ring in lycopene structure (Figure 2). Although its similar chemical formula, the metabolism of lycopene not been studied broadly compared to β-carotene (58)

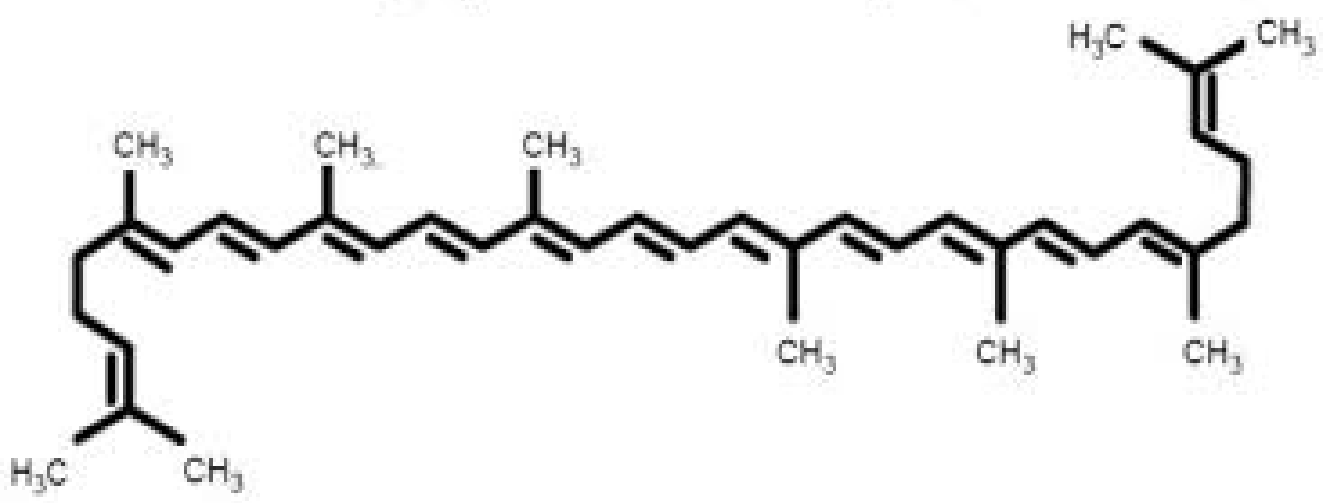

**Figure-1: Structure of Lycopene** (59)

It is alarming that only 10-30% of the dietary lycopene is absorbed through the intestinal lumen due to the unavailability of fats and oils and interaction with dietary fiber, and other carotenoids (60). However, the absorption of lycopene is affected by the adverse processing methods and the presence of lipids and other lipid-soluble compounds. Instead of its low bioavailability, lycopene has twice more significant free radical scavenging potential than pro vitamin-A (β-carotene) and ten times more than vitamin-E (61) which could be the central insight for the lycopene to be selected extensively as an ingredient for functional food development in future.

Jiang et al. (62) reported that if lycopene is consumed it might induce an anti-carcinogenic effect against prostate cancer. According to Jiang, with lycopene treatment, inflammatory factors levels were reduced significantly, such as tumor necrosis factor-α (TNF-α), interleukin1 (IL1), IL6 and IL8. The survival of mice bearing xenografts prostate cancer was significantly improved ($P < 0.01$) by the higher dose of lycopene. However, the tumors' burden can also significantly be reduced ($P < 0.01$) due to lycopene treatment. Lycopene treatment enhances cytoprotective enzymes to be synthesized. Cytoprotective enzymes further could inhibit adenosine deaminase, an essential component for tumor generation. Parveen et al. (171) observed that consuming tomatoes, the richest source of lycopene, can reduce the chance of obesity, hypercholesterolemia, cardiovascular disorder, and different types of cancers. So, lycopene has a bright future for developing functional foods in the food industries (63, 64, 65, 66, 67).

**3. Digestion, absorption, and distribution of Lycopene in human metabolism**

Food containing lycopene ingested through the mouth and starts disintegration through the action of teeth called mastication. Mastication, peristalsis and ptyalin enzymes facilitate the release of lycopene which usually bind tightly by pectin inside the fruits and vegetables cell matrix (68, 69). Ptyalin, secreted from the salivary gland, breaking the pectin structure partially into the mouth and releases lycopene from the cell matrix of fruits and vegetables. Disintegrated foods mix with saliva into the mouth to form bolus which facilitates the transfer of food particles from mouth to stomach (68). As only carbohydrate breaking enzymes were present in the mouth so no degradation occured for the functional properties of lycopene before stomach digestion. When the bolus reaches the stomach, lycopene is released from the food matrix through the enzymatic action of the stomach and HCl secreted from the parietal or oxyntic cells of the stomach wall. In the stomach, lycopene incorporates lipids droplets and forms the lipid micelles which protect the isomerization or degradation of lycopene that happens due to the action of HCl in the stomach. Lipase, secreted from the gastric chief cell in the fundic mucosa of the stomach, also causes degradation of free lycopene into the stomach. Lipid micelles incorporated with lycopene then transfer to the intestine and bind to intestinal enzymes and bile acids for further breakdown before absorption (68). Research evidence that most of the lipid and lycopene breaking down into its monomer in the intestine. Lycopene is absorbed through passive diffusion using the scavenger class B type 1 (SR-B1) (Figure 2) which is also responsible for enterocytic absorption of carotenoids such as β-carotene and lutein (3). However, lycopene breaking enzymes BCO1 and BCO2 partially degrade lycopene in the intestinal enterocyte and reduce the bioavailability of lycopene (6). Lycopene is primarily absorbed by binding with chylomicrons or microsomal triglyceride transfer protein before being transported to the lymphatic system (70, 5). Enterocytes release lycopene from the chylomicrons into the lymph and then to portal circulation (Figure 3) by the action of extrahepatic lipoprotein lipases and clear it to the liver. Low density lipoprotein is the main carrier which receives lycopene from the chylomicrons and transports it to the liver through the bloodstream for the targeted delivery of different organs (71). Target cells are taken up the lycopene through the interaction with SRB1 and CD36 (Cluster of Differentiation 66) membrane proteins (72, Figure 2).

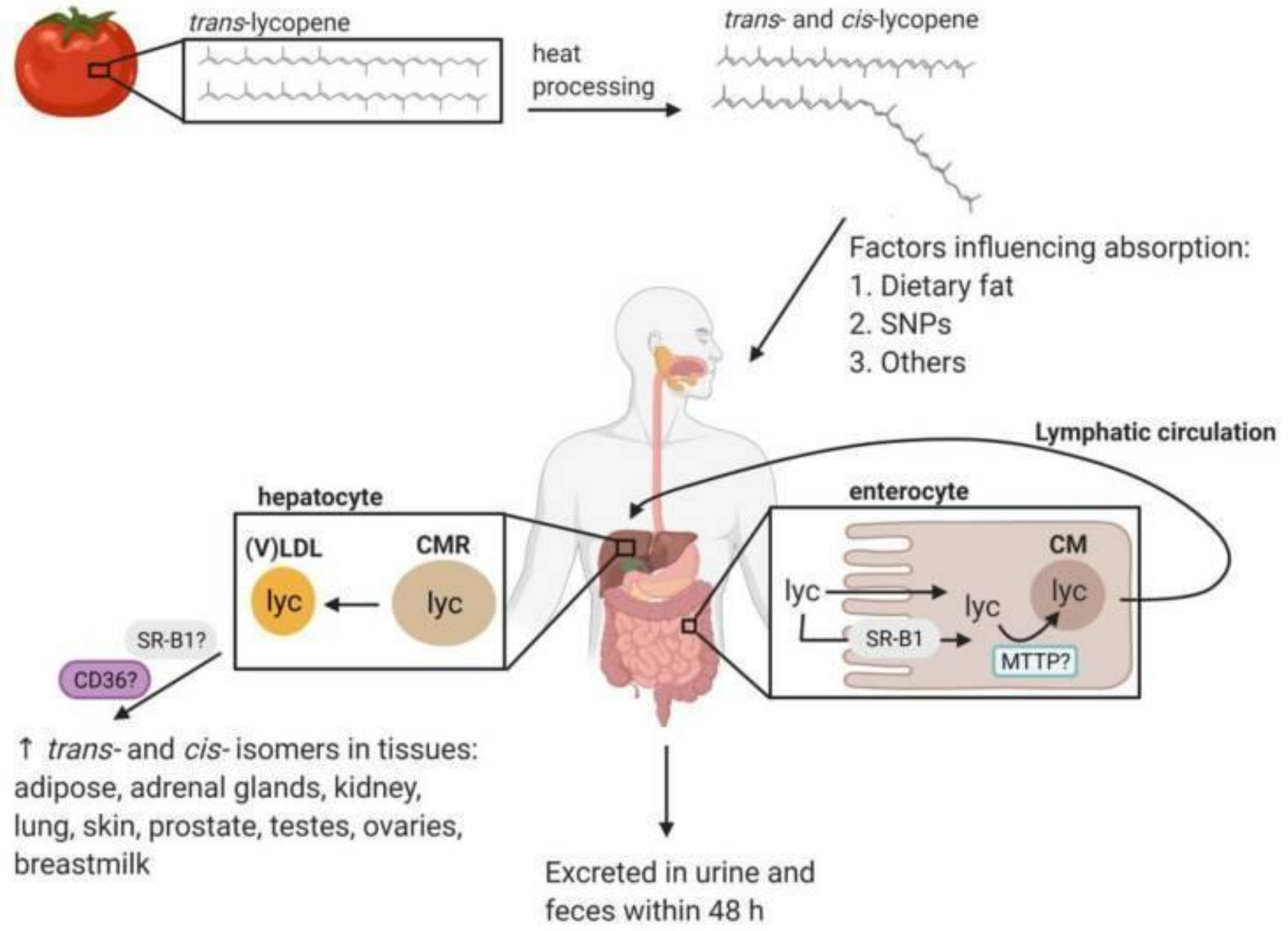

Figure 3. Digestion, absorption and distribution of lycopene through the human metabolic system. Adapted from Arballo et al. (1).

**4. Nanoencapsulation and nanoparticles:**

Nanotechnology has been considered an excellent tool to build a bridge between food and biomedical science, dealing with the manufacturing and application of materials between 1 and 1000 nm (73). Food industries have revolutionized due to the use of nanotechnology in manufacturing functional food products and corresponding ingredients. Nanotechnology is also used for improving the sensory qualities, processability, shelf life, and exclusively the functionality and efficacy of innovative food products for consumer food markets.

Nanotechnology has grasped great attention in advanced science, which helps to encapsulate the bioactive compounds at the nano-level to improve their bioaccessibility and bioavailability through the human diet. The reduction in particle size to the nanoscale range can increase the surface area as well as subsequent reactivity of the encapsulated nanoparticles (NPs) by many folds due to considerable changes in the electrical, mechanical, and optical properties of the NPs (74). Encapsulation of hydrophobic bioactive compounds at nano-level can significantly improve the water solubility, heat stability and bioavailability of the bioactive compounds if ingested through oral route (75, 76, 77, 78, 79, 80, 81, 82, 83, 84, 85, 86).

PLGA (Polylactic co-glycolic acid) is considered an effective carrier (54) for bioactive compounds as its metabolites are digestible into the human GIT through a controlled breakdown mechanism in the TCA cycle. PLA (Polylactic acid), a biodegradable polymer, also popular in encapsulating bioactive compounds (lycopene) due to its biodegradability, biocompatibility, high hydrophobicity, strong mechanical strength, and control release properties (56). Polylactic co-glycolic acid (PLGA) has a faster degradation rate compared to PLA due to high hygroscopicity of PLGA causes higher hydrolysis during enzymatic digestion, also a prominent tool to encapsulate bioactive compounds. As PLGA are hydrolyzed much faster compared to PLA so PLGA can be used for drug delivery of immediate action. Increasing order of rate of hydrolysis is following a chronology: Poly-L-Lactic acid (PLLA) < Poly -D, L-lactic acid (PDLA) < Polylactic co-glycolic acid (PLGA) (87) which mean if faster release of lycopene is necessary the researcher or manufacturer should use PLGA as a coating material for lycopene. In case of slow delivery manufacturers should go for poly-L-Lactic acid. In the presence of alginates, chitosan can encapsulate and affect absorption of lycopene due to its engineered physicochemical properties suitable for improved absorption (88). Besides the coating capability, chitosan contributes positive charge to the particles, has antimicrobial activity, biocompatibility, and biodegradability in human GIT (88). Instead, low methoxyl pectin, a biodegradable polymer, can formulate a gel-like structure that modifies the encapsulation efficiency and release properties of lycopene *in-vitro* (89). This improved encapsulation efficiency also improves the bioaccessibility and delivery efficiency of lycopene to the target cells. Increased delivery efficiency also helps to cure any infectious diseases as well as chronic diseases faster than that of less encapsulated ones. Gum Arabic with an intricate mixture of arabinogalactan and protein grants optical emulsion properties to the lycopene after encapsulation (90). Finally, biopolymers can intensely affect both stability of the encapsulated lycopene against enzymatic degradation, thermal treatments, and auto oxidant and thus improve the bioaccessibility and bioavailability of lycopene if ingested through oral route (91). A thorough investigation on nanoencapsulation of specific bioactive compounds like lycopene and their effect after incorporating them into novel functional foods should be done to satisfy consumer demands and delivery efficiency of health benefits to the consumers. Khathuriya et al. (92), encapsulate lycopene at nano level with gelatin for targeting estrogen receptor of positive breast cancer MCF-7 cells. The stability of the nanoparticles or zeta potential value was 42.1±0.5 mV in case of gelatin alone but 31.1±0.8 mV when gelatin incorporated with estrogen. The drug loading capacity was the same for both lycopene-gelatin (9.16 mg/10mg) and lycopene-estrogen-gelatin nanoparticles (9.22 mg/10mg). Drug release capacity for lycopene-gelatin was 95.0% where lycopene-estrogen-gelatin observed to be 85.7% means that estrogen reduces the drug release capacity of the composite. Bioaccessibility of an encapsulated nanoparticle depends on the physicochemical properties of nanoparticles such as encapsulation efficiency, zeta potential or stability and size of the nanoparticles. So, optimization of physicochemical properties of nanoparticles should be the prime target for improving the bioavailability and targeted delivery of lycopene through oral delivery.

**5. Increase bioavailability or bio-accessibility of lycopene by nanoencapsulation:**

According to Sharma et al. (93), the encapsulation of lycopene by biodegradable polymer revealed a potential rise in blood plasma levels. The relative bioavailability was increased by 297.2% in their experimental lycopene formula than the available marketed formula. This study also inferred better-controlled release, high encapsulation efficiency and better permeability of loaded lycopene into the human intestinal lumen. This study also depicted that zeta potential values did not change in different pH (4 to 9) and temperatures (4$^{o}$C and 25$^{o}$C), where encapsulation efficiency was found to be 62.8±2% at optimized conditions (93).

Sambandam et al. (57) found successful quercetin encapsulation in polylactic acid (PLA), where the hydrodynamic diameter and zeta potential were found to be 220±30 nm and -21.5 ±2.2 mV, respectively. In the same study, encapsulation efficiency was 73.3% in most optimized compositions, contributing to the cumulative release of >99.7% in in-vitro digestion. However, a better scavenging effect was found in encapsulated quercetin than in non-encapsulated one (57). Sanna et al. (94) found PLGA (Poly Lactic-Co-glycolic Acid) to be less efficient coating materials for resveratrol in the presence of chitosan (32%) and alginates (23%). Kassama and Misir (54) found greater encapsulation efficiency for PLGA to encapsulate freeze-dried Aloe Vera gel (86.3%) and liquid (67%) with zeta potential values of -28 and -21.9 mV, respectively, which had better antioxidant properties than the unencapsulated one after the in-vitro human digestion. Therefore, it can be inferred that improving the physicochemical properties can also improve the bioaccessibility or bioavailability of the nano-encapsulated bioactive compounds. Khathuriya et al. (92), observed increase encapsulation efficiency of lycopene improved delivery efficiency up to 95% when gelatin was used as an encapsulating material compared to nonencapsulated lycopene where the absorption range is between 10 and 30%. Dos santos et al. (95) used interfacial deposition method to encapsulate lycopene where PCL used as a carrier material and got an encapsulation efficiency, size and zeta potential of 95.12%, 193 nm and -11.5 mV, respectively. Nazemiyeh et al. (96) used hot homogenization method to encapsulate lycopene and got an encapsulation efficiency of 98.4%. Li et al. (20) encapsulated lycopene by nanoprecipitation method and got the highest encapsulation efficiency of 89%. Singh et al. (97) and Zhao et al. (98) got an encapsulation efficiency of 85.5 and 88.71%, respectively for lycopene. All the researchers observed a significant relationship between improving the physicochemical properties and also improving the bioaccessibility or bioavailability of encapsulated lycopene.

## 6. Formulation of Food grade nanometric delivery systems:

The nutritional value of the food can be improved by adding common bioactive compounds like probiotics, polyunsaturated fatty acids and natural antioxidants, which further improve overall health and wellbeing by preventing the progression of chronic diseases (99). Nanodelivery is a technology that can help to improve solubility (80), functionality (100), cellular uptake (101, 102, 103, 104, 105, 106, 107, 108, 109, 110), stability (111), bioavailability (112) and offer improved controlled delivery efficacy (113) of the biomaterials used in it. Consumer safety has become a foremost concern due to the growing interest in NPs for food and oral drug delivery systems (114). Translocating the NPs into the tissue due to their nano size and higher than physiological dose could be a potential safety concern for the future of this technology.

### 6.1. Type of delivery systems

Nano-delivery systems are divided into two main classes, liquid and solid. The three types of liquid nano-delivery systems are available in the market: nano-emulsions, nanoliposomes, and nano-polymersomes. The most common three types of solid nano-delivery systems are polymeric NPs, nanocrystals, and lipid NPs (5, 95, 115). The lipid particles can also be divided into solid lipid NPs (SLNs) and nanostructured lipid carriers (NLCs). Polymeric NPs can be classified into nanospheres and nano-encapsulates (5, 95, 115). Each of the delivery systems explained is designated to address specific biomedical applications.

#### 6.1.1. Nanoemulsion

Nanoemulsion (Figure 4A) is a system consisting of a mixture of two non-miscible liquids, generally oil and water and dealing with 10 to 100 nm in diameter of droplets. Physically the nanoemulsion is transparent due to its smaller size than the UV light range. Moreover, phospholipids, amphiphilic proteins, or polysaccharides are some common surfactants that stabilize the emulsion (116) and help to develop controlled release, high encapsulation efficiency, and higher protection from environmental degradation (117, 118). In nano delivery, the hydrophobic food bioactive is dissolved in the organic phase of an oil-in-water emulsion, whereas double emulsions are employed to deliver hydrophilic molecules (119).

Two techniques are frequently used to synthesize the NPs, chemical or mechanical processes. Mechanical procedures are highly energy-consuming processes that employ sonicators and microfluidizer to provide larger emulsion droplets into smaller ones. On the other hand, natural development of nanoemulsion droplets due to the hydrophobic effect of lipophilic molecules is considered a low-energy or chemical method. The low-energy method uses emulsifiers to facilitate the nanoemulsion formulation (118). This type of emulsion is usually used in meat industries to replace animal fat from the meat products (120). Additionally, it also used to develop food and beverage flavoring by adding lemon oil emulsion in it (121). But stability of this type of emulsion still has a problem due to instability of the natural emulsifier used for this nanosystem (122, 123).

#### 6.1.2. Liposomes

Phospholipids formed as bilayers provide a major effect on the formulation of liposomes (Figure 4B) self-assembled NPs. Liposomes are standard for holding hydrophilic molecules into the vesicle due to the polar environment in it (124), but liposomes can also help to deliver hydrophobic molecules by encapsulating it into the bilayer. Moreover, liposomes can encapsulate water soluble compounds into its inner layer and fat-soluble compounds in its outer layer. The lipid bilayer protects the encapsulated bioactive compounds from extrinsic conditions (heat, light and oxygen) of food materials. The best part of this type of emulsion is it can formulate nanoemulsion without the use of organic solvents (125) as the phospholipids used to produce liposomes work as an emulsifier. Liposomes can be classified based on their vesicle size. Liposomes deal with the size of the vesicles up to 100 nm called unilamellar, whereas multilamellar vesicles deal within the range between 500 nm and 5 μm. Electrostatic accumulation (125) and hydration (126) are the two most renowned methods to integrate liposomes. Liposomes can be used to make iron-enriched milk (127), or to deliver antioxidants to the human diet (125, 128), and for delivery of vitamins E and C by fortifying it with orange juice (124). However, liposome is not suitable for oral delivery as it is sensitive to stomach HCl dependent hydrolysis and enzymatic degradation due to the action of phospholipases, cholesterol esterase and pancreatic lipase (122, 123). But it is very much effective for drug delivery where immediate action is needed.

### 6.1.3. Polymersomes

Like liposomes polymersomes, which encapsulate both hydrophilic and hydrophobic bioactive compounds, are vesicles made off with amphiphilic copolymer where the bioactive is entrapped by making a bilayer all around it (129, 130). Better control release and higher stability are the main characteristic properties of polymersomes compared to liposomes (129). Moreover, polymersomes are stable under stomach low pH and show sustained release of bioactive compounds such as drugs, oligonucleotides, enzymes and peptides (131, 132). Because of its non-phagocytic or "Stealth" nature (e.g., Poly-ethyl glycol) it can prevent macrophage activation and helps maximum targeted delivery of encapsulated nutrients (131, 132). Due to its stealth nature against macrophage activation is the major drawback for this nano system prepared from biodegradable polymers like lactides and glycolides with the probability of producing potential toxicity if circulate longer time into the extracellular fluids. Extrusion, freezing, thawing cycles and sonication can modify the structure of polymersomes and thus influence the bioaccessibility of encapsulated bioactive compounds and its control release kinetics (133, 134). So, processing conditions must be considered before making a functional food using polymersomes.

### 6.1.4. Nanocrystals

In nanocrystals (Figure 4D), poorly water-soluble bioactive ingredients are enclosed by a surfactant to enhance their solubility in-vivo (135, 136). The encapsulation efficiency of bioactive ingredients in nanocrystals is almost 100% which can be formulated without using organic solvents. There are two ways to formulate nanocrystals, mechanical and chemical techniques (135). Acid hydrolysis and ultrasonication are chemical and mechanical methods used prominently to form polymeric nanocrystals, respectively. Continuous stirring should be ensured to inhibit the nanocrystals from aggregation (137, 138). Water is commonly used to increase the homogeneity and reduce the particle size of the crystal (135).

The nanoprecipitation method is considered the faster and more efficient method to produce nanocrystal (135). Moreover, carbohydrate-based polymers and protein are very popular in manufacturing nanocrystal based novel functional foods for biomedical applications (139, 140, 141) but are not suitable for biomedical application through oral delivery if encapsulating bioactive compounds are sensitive to enzymatic degradation. The reason behind this unsuitability can be verified by enzymatic activity of saliva, gastric and intestinal juices. Ptyalin from saliva and pancreatic amylase can break down carbohydrate into short chain polymer dextrose where gastric and intestinal protease degrade protein into amino acid and releases encapsulating bioactive compounds to the GIT. After that, when enzyme sensitive bioactive compounds are exposed to saliva, gastric & intestinal juices it reduces its bioaccessibility by oxidation and isomerization type of enzymatic reactions due to the activity of amylases, lipases and proteases (142).

### 6.1.5. Lipid Nanoparticles

Like nanoemulsions, lipids are used to make lipid NPs (SLNs) but in a solid-state. The significance of the solid lipids used to produce NPs is their slow enzymatic digestion and contributes to controlled release of the bioactive compounds embedded in them (118). The organic solvent is needed to formulate lipid NPs where lipids work as a coating material for bioactive compounds (143). Lipid NPs used for releasing those drugs need immediate release into the human GIT to perform their work quickly (124).

Ultra-Sonication and microfluidization consuming higher amounts of energy are popularly applied to synthesize solid lipid NPs for food applications (124, 118, 143). Surfactant concentration is the most critical factor for SLNs formulation. Researchers should be cautious in using surfactant concentration as too high or too low of surfactant contributes to the aggregation of SLNs (Figure 4C). Multiple surfactant composite can accelerate the stability of the emulsion where ionic surfactant contributes smaller particle sizes (143). Scientists are using SLNs (Figure 4C) to increase the shelf life of fruits like guava (144), the constancy of antioxidants (145), and prominent carotenoids like β-carotene and vitamin-E (111). Stability of lipid nanoparticles is very low which can be improved by the combination of more than one surfactant into the system (143).

### 6.1.6. Polymeric Nanoparticles

When polymers are used to encapsulate bioactive compounds in the presence of surfactants are known as polymeric NPs (146, 136, 42) (Figure 4E). The primary significance of these particles is to protect the encapsulated bioactive ingredients from external environmental factors and enzymatic digestion (147).

Polymeric NPs can be synthesized by solvent nanoprecipitation, emulsion evaporation, and salting out techniques (148). Furthermore, the engineering of carbohydrate-based polymeric NPs can extensively improve the bioaccessibility and bioavailability of bioactive ingredients (149). It also improves the control delivery of bioactive compounds for a longer period of time. Lack of toxicity and safety information could be a potential problem to be used for nano drug delivery systems (150).

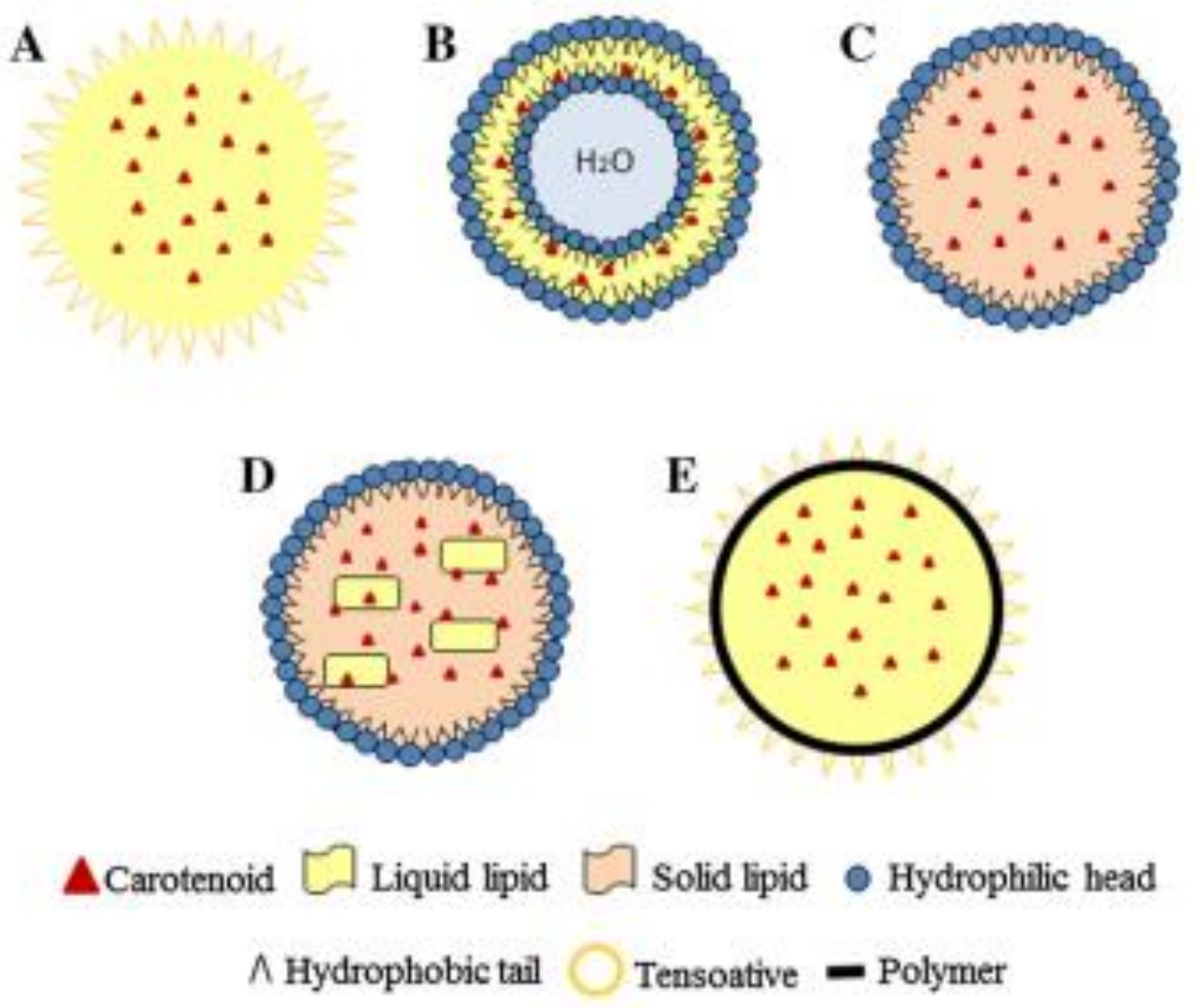

Figure 4. Type of nano-delivery system used to encapsulate lycopene. A = nanoemulsion, B = Nanoliposomes, C = Solid lipid nanoparticles, D = Nanostructured lipid carrier, E = Polymeric nanoparticles. Adapted from dos Santos et al. (2018) (151)

## 7. Factors to be considered for nanometric delivery system:

The nanoparticles' physicochemical properties affect their absorption, distribution, metabolism, and excretion (ADME), which is essential in modulating the *in vivo* delivery of bioactive compounds. Major physicochemical attributes of NPs conducive to *in vivo* interactions are size or hydrodynamic diameter, charge or zeta potential value, hydrophobicity, and target delivery kinetics.

### 7.1. Size of the nanoparticles:

The size of the NPs is the prime most important determinant for intestinal absorption, stimulation of immune cells, and excretion through the large intestine (152). Nanoparticles usually enter the cells through endocytosis. After that, NPs enters the cell and establishes the translocation of the encapsulated biomolecule contained by the endosome (153). Particles sized between 60 nm and 1,000 nm commonly go through the endosome. However, enterocytes prevent the endocytosis of the particles larger than 500 nm and therefore excreted through the feces before entering the bloodstream (109). Moreover, the size of particles can affect the immune sensitivity as the particles size larger than 200 nm are more likely to be eradicated by the mononuclear phagocytic cells of the bloodstream and thus helps to remove potential foreign particles which would cause cell damage (154). Particles, ranging from 30 to 100 nm, can escape the mononuclear phagocytic cells and prevent clearance from the bloodstream, allowing for long circulation times and enhanced targeted distribution (155). Based on the previous evidence it can be said that size of the nanoparticles should be maintained below 200 nm or optimally between 30 to 100 nm for making a potential targeted delivery system of bioactive compounds encapsulated at nano-level.

### 7.2. Hydrophobicity:

Hydrophobicity affects cellular absorption, targeted delivery, interaction with plasma proteins, immune cells and, and excreted from the GIT (152). Hydrophobic NPs diffuse into epithelial cells more effectively than hydrophilic NPs because of a larger activity coefficient (156, 157). However, hydrophobic NPs are identified as foreign elements by macrophages of the mononuclear phagocyte system in the human blood and are frequently excreted through the urine (152, 158). But for macrophages of the mononuclear phagocyte to be active nanoparticle size should be above 200 nm (154). So, if nanoparticles' nature can be hydrophobic and size is maintained below 200 nm then a better delivery system can be developed to address chronic diseases through oral delivery of encapsulated bioactive compounds.

### 7.3. Hydrophilicity:

Hydrophilicity affects the extracellular excretion of nanoparticles. Hydrophilic NPs quickly excreted through the kidney (154) due to their smaller size (<10 nm) and less likely to be identified as hazardous materials in the bloodstream (152). In case of its faster excretion nanoparticles are not able to release its core bioactive compounds into the bloodstream for its targeted delivery and thus losses its bioavailability due to excretion through the urine. In addition, the hydrophobic interaction of NPs prevents the navigation through mucus and provides slower excretion rate than that of hydrophilic particles and get enough time to release bioactive specially from polymeric core of NPs (159). It can be inferred that increasing hydrophilicity of the nanoparticles decreases the absorption capacity of encapsulated bioactive compounds. So, for providing maximum bio-accessibility of the nano-encapsulated bioactive compounds, hydrophobic nanoparticles should be taken as optimum conditions.

### 7.4. Charges of the nanoparticles:

Charge of the NPs significantly affects the absorption through the intestinal cells, interaction with immune cells & plasma protein and contributes toxicity production (152). Positive charges of the NPs are more susceptible to improve an immune response than negative and neutral charges (152, 160). Moreover, NPs without charge remain in the bloodstream longer than positively or negatively charged particles (158) and thus targeted delivery of NPs affects distribution and immune response in human physiology (161, 152). In addition, the physicochemical properties like size and surface charge affect the target delivery of molecules, which further affect immune response in the human metabolic system (161).

Table-1 shows that multi-dimensional factors are responsible for modifying the physicochemical properties of lycopene NPs. Ha et al., 2015 found that high pressure homogenization method facilitates the NPs formulation where higher the pressure lowers the hydrodynamic diameter and stability or zeta potential value of the NPs (196). Unlikely, encapsulation efficiency of lycopene decreases with increases of homogenization pressure. Li et al. (20) and Ha et al. (196) used nanoprecipitation technique to encapsulate lycopene

where higher the lycopene concentration higher the hydrodynamic diameter but lower the zeta potential and encapsulation efficiency. “Rotary evaporation film ultrasonication” method giving a small NPs size for lycopene ranges from 58 to 105 nm in diameter but high zeta potential value or stability of the NPs ranges from -37 to -32.5 mV. Hot homogenization methods providing more encapsulation efficiency for lycopene ranging from 86.6±0.06 to 98.4±0.5% (96). Interfacial deposition method also depicting high encapsulation efficiency (95.12±0.42%) for lycopene NPs synthesized in PCL (Polycaprolactone) (44). High pressure homogenization method provides highest stability for encapsulated lycopene NPs (ranges from -74.2 to -74.6 mV) and lowest polydispersity index (0.13±0.02 to 0.15±0.05) value (162). In a nutshell it can be said that high pressure homogenization is the best method for high encapsulation efficiency, low polydispersity index and high stability which are prominent characteristics for high bioaccessibility and improved targeted delivery. Hot homogenization is also a potential alternative of high-pressure homogenization which is giving high encapsulation efficiency for lycopene instead of better stability as high pressure is incorporated with high production cost due to high energy consumption.

## 8. Challenges toward nanodelivery of bioactive compounds through functional foods

Nano encapsulated bioactive ingredients can be used to fortify different juices or drinks for delivering bioactive compounds to human nutrition. Nevertheless, some potential problems should be addressed before using the nanoparticles as a functional ingredient for fruit juices and making them available for the people in the consumer market:

a. Fortification can improve the pH, acidity, color, viscosity, taste, and mouthfeel of the juices, which would impact the overall consumers' behavior toward the products. Though nanoencapsulation provides potential protection against the release of bioactive compounds, the pH, acidity, and interaction of other nutrients would release a certain degree, which can negatively affect the sensory properties of fortified liquid juices.
b. Even if the encapsulated bioactive has no undesirable effect on the properties of the fortified fruit juices, the producer must deal with some strict regulation which at some point is not even clear to the producer or even not in place yet.
c. The toxicity of the nanoparticles at specific concentrations is still not understood well.

Long term toxicity study is needed to evaluate the actual effect of newly nanoparticles. Therefore, some extensive scientific research is required to verify the guidelines around the safety levels of encapsulated bioactive ingredients. Digestion and absorption profiles are crucial for identifying the actual delivery efficiency of bioactive compounds to the target cell, which ultimately safeguards the development of marketable nanoparticles-based functional foods.

## 9. Toxicity profile of polymeric nanoparticles:

Toxicity is a major concern for polymeric nanoparticles to be used for biomedical applications. Because of the lack of toxicity data of polymeric nanoparticles, liposomes gain popularity for biomedical applications. A comparative discussion is given below for lipid and polymeric based nanoparticles

## 9.1. Lipid-Based Nanoparticles

Liposomes, composed of phospholipid layer, bearing hydrophilic head and hydrophobic tails which helps it to act as an amphipathic nanocarrier for a wide range of drug therapies (163, 164). Considering the toxic effect and control delivery of liposomes about more than 40 liposome-based drugs have already been approved or under clinical trial for pharmaceutical market (165, 166). These liposomes-based drugs usually interact with LDL, HDL and opsonins which further enhance activation of reticuloendothelial system to eliminate it from the metabolic system (167). FDA approved PEGylated liposome to avoid opsonization and excretion by phagocytic cells (165). However, Gabizon et al. (168) identified an encapsulation technique for a type of chemotherapy drug called an anthracycline (doxorubicin) that can alter its excretion profile in human metabolism. FDA has approved two lipid nanoparticle-based vaccines of PfizerBioNTech with 95% efficiency very recently (169, 170). Modification of physicochemical properties of lipids nanoparticles is necessary to retain these vaccine's stability (171). Alnylam Pharmaceuticals developed a lipid RNAi-based drug called ONPATTRO working against diseases initiated because of altered Transthyretin protein also approved by FDA EC in 2018 (171). Supplementation of vitamin A and premedication of corticosteroids and antihistamines with liposomes come up with side effects like blood pressure, headache, and respiratory symptoms. Despite extensive effort has been made to implement nanotechnology in chronic and fatal disease recovery, a few data are available for liposomes to be used for pregnant and renal patients (166, 11).

Complement activation, hypersensitivity reactions (HSRs), cardiopulmonary distress, and anaphylactoid reactions are anchored due to negative charge of PEG-liposome (172). Liposome size, chain length and surface concentration have mild effect but when PEG incorporated with cholesterol appeared maximum complement activation (172).

## 9.2. Polymeric Nanoparticles

Despite limited data on toxicity, biocompatibility, and physiological issues, polymeric nanodelivery of drugs has become popular for therapeutic treatments (173). Polymeric NPs' toxicity is affected by quantum size effects, which are linked to oxidative stress, cytotoxicity, and genotoxicity (174). Cytotoxicity on human-like THP-1 macrophages appeared when PLGA nanoparticles were incorporated with chitosan for its stabilization (171). Similarly, poly vinyl alcohol and poloxamer 188 polymers are also evidenced to be cytotoxic when incorporated to PLGA as a stabilizer during its nanoparticle formulation. These inferred that PLGA nanoparticles appeared toxicity only if stabilizers were used to manufacture the nanoparticles (2).

Cationic stearyl amine PEG-PLA nanoparticles observed higher local and systemic harmful effects on mice model (175). However, replacement of synthetic nanoparticles and organic solvents with natural polymers can potentially alleviate the toxicity effect of encapsulated nanoparticles raised on mice model (176). The main challenges for polymeric nanoparticles to be used for chemotherapeutic drug delivery are their short clearance time and poor targeted delivery. Palanikumar et al. (177), used a cross-linked bovine serum albumin shell to the drug loading system to preserve it from serum proteins interaction and macrophage dependent clearance and thus improve clearance and targeted delivery of liposomes. As a result, the drug-loaded NPs proved as a strong anticancer agent both *in vitro* and *in vivo* without producing any toxicity when the serum cross-linking technique is applied on NPs formulation system (177).

**10. Release kinetics of encapsulated bioactive compounds:**

Release mechanisms of encapsulated bioactive compounds depend on the composition of the particles, stability or the charge of the NPs, preparation method and release media. Release kinetics can be explained by mathematical popular models based on their phenomena and predictive efficiency (178). Majority of the time single models are not capable of explaining a drug delivery system where multiple models can explain very efficiently and precisely. Encapsulated polymer solubility, dose of the bioactive, molecular weight and size of the polymer, NPs size and shape are some potential influencing factors for explaining the release kinetic of encapsulated bioactive compounds by different mathematical modeling (179). Based on the literature review only those models are described below which were considered for polymeric delivery.

**10.1. Higuchi and Baker & Lonsdale Model**

The first example of a mathematical model intended to describe bioactive ingredients released from an encapsulated core matrix is called Higuchi model (180). The Higuchi model is majorly valid to identify the release of water soluble and low soluble drugs incorporated in gel and solid polymeric matrices. The model expression is given by the equation:

$$Q = A\,[D\,(2C - C_s)\,C_s\,t]^{1/2} \qquad (1)$$

where Q is the amount of drug released in time t per unit area A, C = initial concentration of drug, $C_s$ = solubility of drug in the media and D = diffusion coefficient of drug.

Simplified Higuchi model describes the release of drugs from an insoluble matrix as a square root of the time dependent process based on the Fickian diffusion equation.

$$Q = KH\,t^{1/2} \qquad (2)$$

The equation can be explored by plotting the cumulative percentage drug release value against the square root of time. The slope of the plot is the Higuchi dissolution constant (KH). The major benefit of this equation is that it can include the possibility to facilitate device optimization and explain the underlying drug release mechanisms.

Baker and Lonsdale (1974) (181) developed a model with the improvement of the Higuchi model to describe drug release from spherical shaped materials which can be explained by following equation:

$$f = 3/2\,[1- (1-Mt/M\alpha)^{2/3}] - Mt/M\alpha = Kt \qquad (3)$$

where Mt / Mα is the fraction of drug released at time t and for simplified appearance it can be denoted as Q and the equation will be f = 3/2 [1- (1-Q) 2/3] – Q = Kt. The slope of the plot will be the release constant (K) (233).

From equation (1) depicted that solubility of a drug is a major component for it. As lycopene is insoluble in water, the Higuchi model is not the appropriate model to explain lycopene release kinetics. On the other hand Baker & Lonsdale model the prime criteria are to be the spherical shape of encapsulated bioactive and the release constant which also depends on water solubility of bioactive which is zero in case of lycopene. So, It is not a suitable model for explaining the lycopene release kinetics.

**10.2. Hixson-Crowell cube root law**

Hixson and Crowell (182) first proposed a cube root law to represent dissolution or decreasing rate of solid surface area as a function of time. This model is mostly applicable for a system where the erosion happens on surface area and hydrodynamic diameter of particles with time. This cube root law can be written as

$$Q_t^{1/3} = Q_0^{1/3} - K_{HC}t$$
$$Q_0^{1/3} - Q_t^{1/3} = K_{HC}t \quad (4)$$

where $Q_t$ = remaining weight of solid at time t, $Q_0$ = initial weight of solid at time t = 0, and $K_{HC}$ represents the dissolution rate constant.

Every polymeric encapsulated nanoparticle releases its bioactive form through erosion on surface area and hydrodynamic diameter of particles with time. But cube root law preferentially restricts the release kinetics into a certain frame which also compromise the universality of lycopene release from different polymeric nanoparticle cores as different polymers have different degradation profiles for the same encapsulated bioactive like lycopene. This model did not consider hygroscopicity or swelling capacity and burst release kinetics of polymer. It is also not capable of identifying the type of release kinetics (e.g. is it Fickian or non-Fickian) appearing into the system which would have a great impact in the targeted delivery system of lycopene for chronic diseases prevention.

**10.3. Korsmeyer-Peppas Model**

Korsmeyer et al (183) formulated a simple relationship which explained the bioactive compound release from a polymeric core-based matrix. Korsmeyer and Peppas (186) and Ritger and Peppas (184, 185) established experimental equations to differentiate between Fickian and non-Fickian release of bioactive compounds from swelling as well as nonswelling polymeric delivery systems. The Korsmeyer Peppas (114) model is explained by following equation:

$$M_t/M_\alpha = kt^n \quad (5)$$

where $M_t$ = amount of drug released, $M_\alpha$ = initial drug load, $Mt/M_\alpha$ = fraction of drug released at time $t$, $k$ = Korsmeyer-Peppas constant, $t$ = time, $n$: release exponent. The release mechanism of the lycopene can be determined based on the release exponent ($n$) value mentioned in table 2 (187).

Korsmeyer Peppas model considered swelling or non-swelling properties of polymers. It also can differentiate Fickian and non-Fickian types of diffusion based on the exponent value of time (t) (equation 5). It did not consider the shape of the particles as variables which means shape does not have any effect on the release mechanism of bioactive compounds encapsulated from the polymeric core. But it's a very good model to explain the release of lycopene from the nanoparticles core if the shapes of all particles are the same.

**10.4. Hopfenberg model**

Hopfenberg (188) developed a mathematical model which explained the release kinetics of bioactive compounds or chemical compounds by surface erosion of polymers until the surface area remains constant during this erosion process. The cumulative fraction of the drug released at time *t* was described as

$$Mt/M_{\alpha} = 1 - [1 - K_0 t/CL\ a]^n \qquad (6)$$

where $K_0$ is the zero-order rate constant describing the polymer degradation (surface erosion) process, CL = initial drug loading throughout the system, a = system's half thickness, and n = exponential value that varies with geometry n = 1, 2 and 3 for slab (flat), cylindrical and spherical geometry, respectively.

Hopfenberg considered the shape of the nanoparticles as a variable for release kinetics of bioactive compounds. Where it cannot differentiate different types of diffusion appeared in releasing bioactive compounds from the nanoparticles core.

**10.6 Gallagher Corrigan model**

Bioactive compound releases from its biodegradable polymeric carrier by the combination of diffusion and degradation mechanism called a biodegradable polymeric drug delivery system. Polymer degradation profiles usually have a sigmoidal in shape which can be elaborated best by Gallagher and Corrigan mathematical model (189). Kinetic profile of polymer degradation described by Gallagher-Corrigan equation consisting of the initial 'burst effect' contributed by bioactive adsorbed on the surface of NPs matrix and following slow subsequent release due to the erosion of polymeric matrix (189). The total fraction of drug released ($f_t$) at time (t) is

$$f_t = f_{tmax}\,(1-e^{-K_1.t}) + (f_{t\,max}-f_B)(e^{K_2.t - K_2.\,t_{2\,max}})/(1 + e^{K_2.t - K_2.t_{2max}}) \qquad (7)$$

where $f_t$ = fraction of drug released in time (t), $f_{tmax}$ = maximum fraction of drug released during process, $f_B$ = fraction of drug released during 1$^{st}$ stage – the burst effect, $k_1$ = first order kinetic constant (1$^{st}$ stage of release), $k_2$ = kinetic constant for 2$^{nd}$ stage of release process–matrix degradation, $t_{2max}$ = time to maximum drug release rate. This calculated $f_t$ is plotted against the time, and the correlation coefficient and coefficient of determination can be calculated to understand the suitability of the model.

Gallagher and Corrigan model (189) is a very complex but elaborative model considering burst release, different stage of release and the time for maximum drug release, are topmost important for drug delivery to prevent chronic diseases, as a variable and make the model most acceptable for explaining the release kinetics of bioactive compounds.

Encapsulation of lycopene can be done by liquid lipid, solid lipid, polysaccharides, and different biodegradable polymers. Control release kinetics are found to be different in literature based on the physicochemical properties of coating materials and type of coating materials. Burst lycopene release found both for polymeric and lipophilic types of particles (Figure-2 (a)). For nanoliposome, 80% of the total lycopene can be released within 12 hours (Figure-2 (a)) of its exposure to the phosphate buffer solutions (190). However, 50% of the total loaded lycopene released within 24 h (Figure-2 (b)) of its contact to the releasing solution (160) when encapsulated with N-isopropylacrylamide where 60% (Figure-2 (c)) is released within 3 h from solid lipid core of the lycopene nanoemulsion (96). For polymeric NPs lycopene releases faster at lower lycopene concentration. PLGA based coatings are capable of releasing more than 80% of the total lycopene by half an hour of contact time with phosphate buffer solution at pH 7.4. Consistent release of lycopene for a longer time is the main challenge of future progress in food nanotechnology. Based on the above description it can be seen that lycopene if encapsulated in lipid, protein or any polysaccharides/polymer burst release appeared at the beginning or at the end of the release mechanism. Even every research article also found multiple stages of release which can only be explained by Gallagher and Corrigan model (189). It can be conferred that the Gallagher and Corrigan model is the best to explain lycopene release kinetics irrespective of the nature of coating materials if it is encapsulated at nano-level.

In case of nanoencapsulation or nanosphere formulation some of the bioactive compounds embed inside the nanocapsule or nanosphere and some just adsorbed on the surface of the nanocapsule or nanosphere (Figure 5). Because of this special orientation of bioactive compounds, bioactive compounds attached on the surface of the nanoparticles release faster at the beginning of release kinetics and initiate early burst release. But the bound bioactive, embedded inside the particles, release kinetics depends on the degradation sensitivity of coating materials and sometimes introduce 2$^{nd}$ burst release when coating materials completely lose their physical strength. Based on this explanation burst release is a natural tendency of nanoencapsulated bioactive compounds and Gallagher and Corrigan model should be the best model to explain this special characteristic of nanoparticles. On the other hand, polymers are hygroscopic, and type of diffusion identification is also an important variable for targeted drug delivery systems which can be elaborated by Korsmeyer Peppas model. So, combination of Korsmeyer Peppas and Gallagher & Corrigan model should better explain the release kinetics of lycopene from polymeric nanoparticles core.

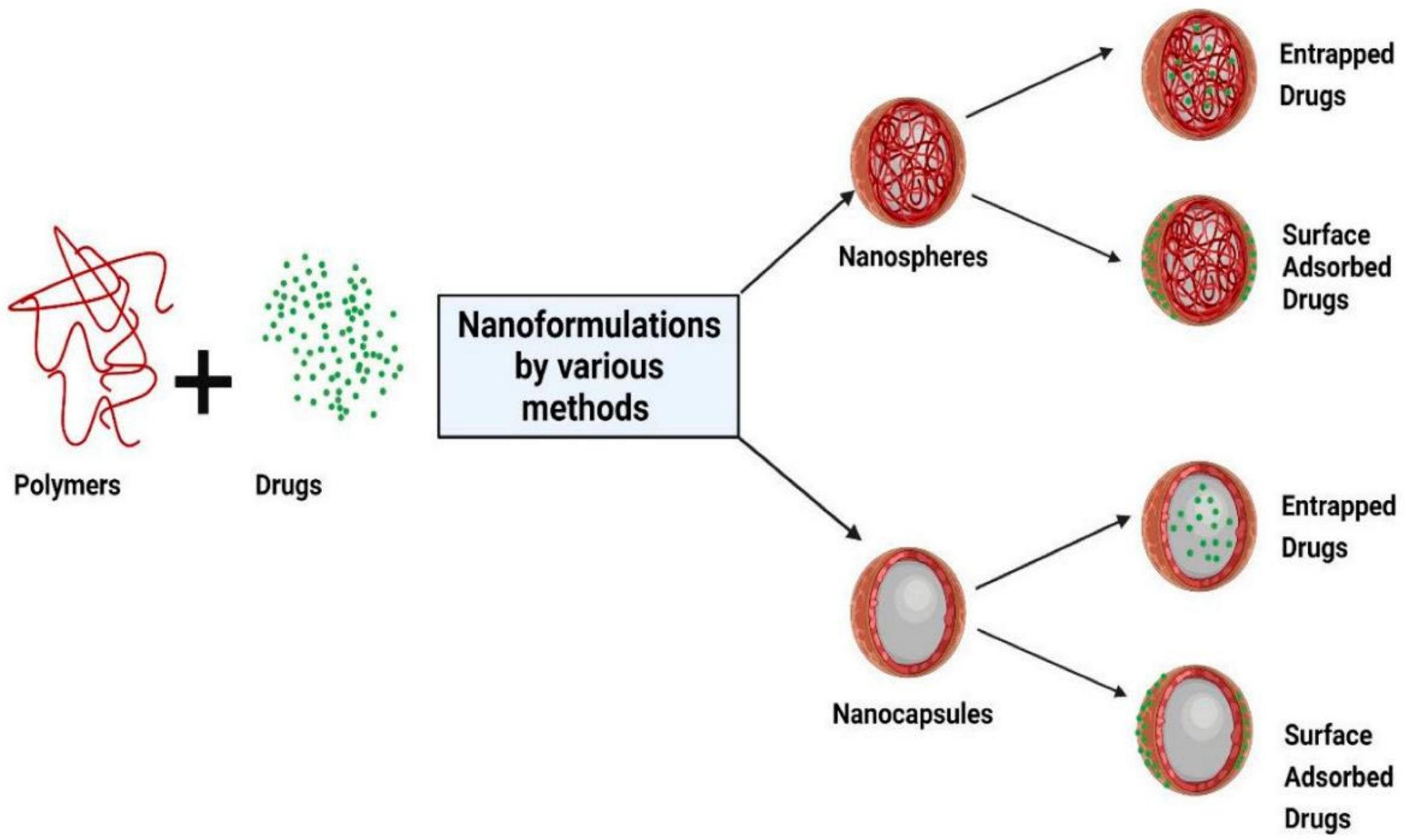

Figure 5: Formulation of polymer base nanocapsule and nanosphere matrices (191)

Controlled and targeted delivery mechanisms of nanoencapsulated carotenoids is very much vital for developing a new drug system. After formulating the encapsulated carotenoids (e.g. beta carotene or lycopene) if it administers into the rat model it should passes through the intestinal lining of the rat model, moving towards the extracellular fluid and penetrate into the cell for control delivery of bioactive compounds in it (Figure 6). If the release kinetics is known, then it would be easy to establish antioxidants and reactive oxygen species balance inside the cell matrix and thus prevent the onset of chronic diseases (Figure 6).

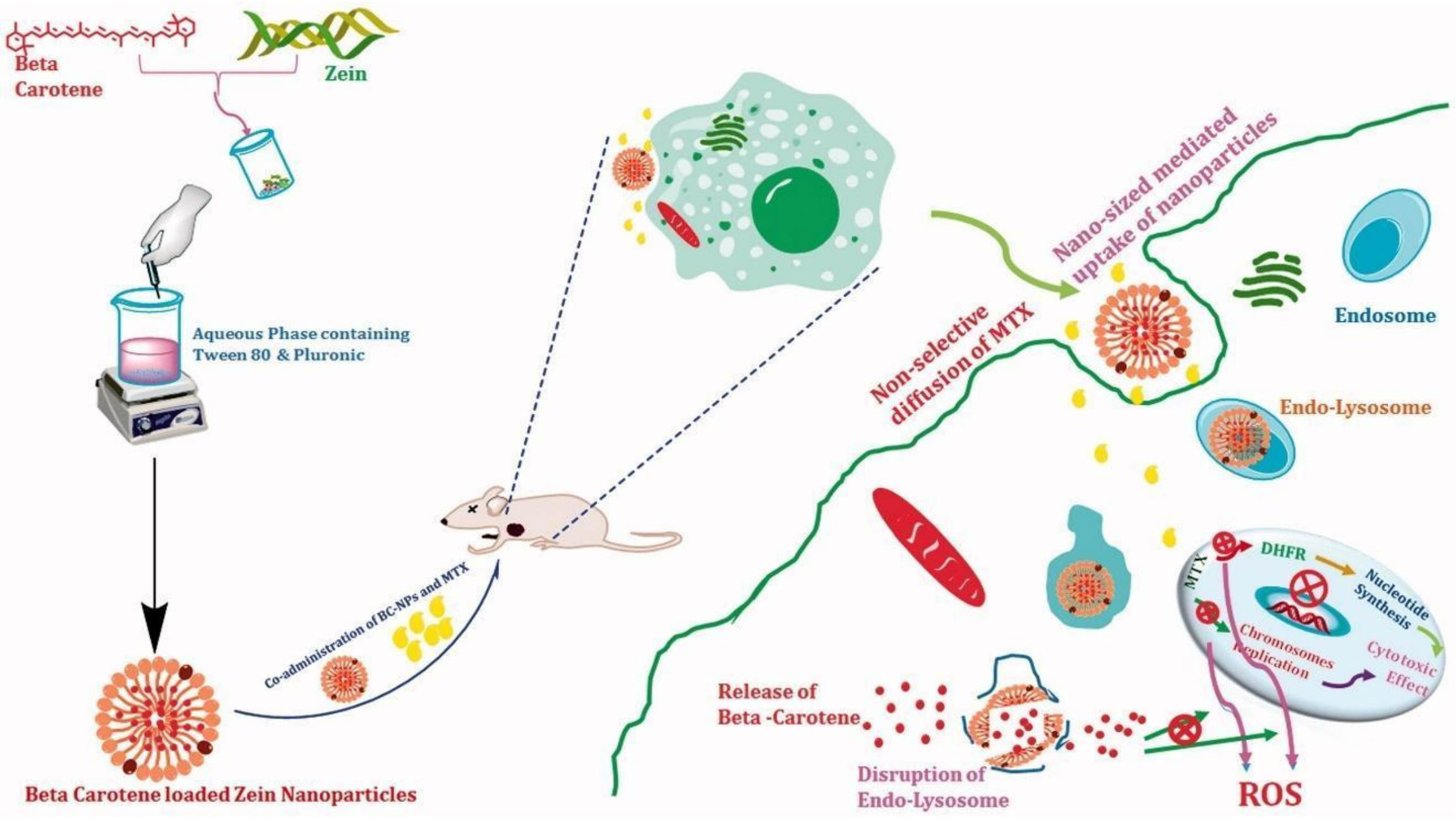

Figure 6: Controlled and targeted delivery of carotenoids (e.g. beta carotene or lycopene) into the cell. Adopted from Jain et al. (192).

## 11. Challenges and Future of Nanotechnology in Food Science

Nanotechnology in food science is a greenfield for disease prevention and treatment through dietary supplementation of powerful antioxidants like lycopene by encapsulating it in a biodegradable polymer. However, due to the high sensitivity of lycopene to oxygen, light, and temperature, the manufacturing technique even has a degradation effect on lycopene bioactivity before pushing it into the polymeric core during the nanoencapsulation procedure. So, most optimized techniques should be identified, and appropriate verification should be followed to predict its degradation profile and its effect on realistic release kinetics during in-vitro human digestion. Researchers have developed mathematical modeling to explain the control release kinetics of the bioactive compounds from the core of polymeric NPs. However, every mathematical model has specific phenomena, which should be well understood for adopting it as a verification tool.

Presently, potential toxicity is the major concerning issue of nanotechnology implementation in biomedical application. By-products of NPs, such as antibody fragments, nucleic acids, peptides, and proteins, can function as antigens, resulting in increased immunotoxicity (193). Moreover, a technique's low encapsulation efficacy causes adverse effects by providing a large amount of coating materials and surfactant instead of contributing bioactive ingredients desired to supply through the diet. Most of the study designs were established to observe short-time toxicity, whereas long-time toxicity remains not to be well understood. Currently, there is no in-vitro model to simulate the human digestion of NPs (194) as well. However, how researchers can cross the biological and physical barriers, such as blood-brain barrier, blood vessel wall, interstitial fluid pressure gradients and extracellular matrix, existing between NPs and abnormal cells is also not known to the researcher which should be overcome to facilitate the delivery of NPs to the target cells (195).

Administration route identification for NPs delivery could be a challenge as the most practical and acceptable route, the oral route, causes degradation due to its enzymatic action on saliva, gastric and intestinal juices. However, little knowledge is available about the absorption and metabolism of NPs in the gastrointestinal tract, thus limiting the idea of a research project about the bioavailability and tissue-specific pharmacokinetics of the nanocarriers.

Finally, the cost generated by this technology is another key issue of its implementation. Indeed, optimization of NPs synthesis, where the preventive and therapeutic bioactive compounds require unique ingredients, specific instruments, and optimal conditions, thus making it expensive (193). Therefore, considering both business and nutrition points of view, lowering the cost/benefit ratio is challenging for implementing nanotechnology in food and biomedical research.

## 12. Conclusion:

Research evidenced that carotenoids are a powerful antioxidant group that helps to prevent chronic diseases if ingested daily. Lycopene has the strongest antioxidant properties compared to other carotenoids due to its 11 conjugated and two unconjugated double bonds. Moreover, processing parameters such as light, heat and oxygen can reduce its bioaccessibility by oxidization and isomerization reaction. On the other hand, due to deficiency of fat in the diet, excess fiber and phytate content and other nutrient interaction pushes down lycopene's absorption to 10-30%, ultimately fading its anticarcinogenic properties if ingested through the oral route. Nanotechnology offers a platform for bioactive compounds like lycopene to improve its bioaccessibility by encapsulating it into its nanocarrier such as fat, protein, polysaccharides or other carbohydrate base polymer. Different nano-delivery systems (liposomes, polymersomes, nanoemulsion, nanocrystal, lipid nanoparticles etc) are established to prevent lycopene degradation to use it as a tool to prevent chronic diseases. Each nano-delivery system exists with its distinct properties and provides a unique delivery mechanism to the bioactive compounds at the biological environment.

Moreover, the polymeric delivery system is the most stable among others and has the potential control over the targeted delivery of lycopene to the cell. A different mathematical model has been developed to explain the release kinetics of each delivery system. Still, none of the single models completely explain the release kinetics of a single delivery system. In the case of a polymeric delivery system, the mathematical model should consider the hygroscopicity and burst release properties, as these are some common phenomena for polymeric delivery systems. Based on these two prime criteria, Korsmeyer Peppas and Gallagher & Corrigan model mutually can explain the delivery mechanism of lycopene from the polymeric core to the targeted cells. Korsmeyer Peppas model considers the polymeric delivery system's hygroscopicity and helps to identify diffusion type based on its time exponent value. On the other hand, Gallagher & Corrigan model explains the burst release emerged due to both adsorbed lycopene on the surface of the polymer at the beginning and pH & enzyme-dependent erosion of the polymer at the end. From a nutritional point of view, nanotechnology offers promising approaches to preventing chronic diseases. But cost-effectiveness and long-term safety study (consecutive *in silico*, *in vitro*, and *in vivo)* of lycopene nanoparticles should be done to get the positive output from the nanoencapsulation procedure before applying it to the food system and biomedical application.

**Appendix**

Table-1: Physicochemical properties and encapsulation efficiency profile of lycopene NPs at different compositional methods

| Method | Composition | Categories | HD (nm) | PDI | ZP (mV) | EE (%) | References |
|---|---|---|---|---|---|---|---|
| Emulsion evaporation+ High | -Ethyl acetate (5%) | Pressure of 60 MPa | 308±57 | 0.35 | 41.6±0.5 | 65.3±1.41 | Ha et al., 2015 [196] |
| pressure homogenization | -Butylated hydroxy toluene (0.01%)<br>-Tween-80 (20) | Pressure of 80 MPa | 290±75 | 0.47 | 41.3±1.2 | 62.55±6.06 | |
| | | Pressure of 100 MPa | 266±97 | 0.38 | 39.9±1.3 | - | |
| | | Pressure of 120 MPa | 221±58 | | 39.2±1.3 | - | |
| | | Pressure of 140 MPa | 184±20 | 0.28 | 33.7±0.2 | 51.63±6.06 | |
| High pressure homogenization | Orange wax + rice oil (9:1) | Lycopene (5 mg) | 158±3 | 0.13±0.02 | -74.6±1.13 | - | Okonogi et al., 2015 (162) |
| | | Lycopene (25 mg) | 160±4 | 0.14±0.03 | -74.2±1.6 | - | |
| | | Lycopene (50 mg) | 166±4 | 0.15±0.05 | -74.6±2.0 | - | |
| Interfacial deposition | PCL (200 mg) Capric/caprylic triglycerides (320 µL) Sorbiton monostearate (76 mg) Polysorbate-80 (154 mg) | - | 193±4.7 | 0.069±0.02 | -11.5±0.4 | 95.12±0.42 | Dos santos et al., 2015 (151) |
| Hot homogenization | Glyceryl palmito stearate (GPS) Myristic acid Poloxamer-407 (PL-407) | GPS (650 mg) MA (90 mg) PL-407 (690 mg) | 125±3.89 | 0.159±0.105 | - | 98.4±0.5 | Nazemiyeh et al., 2016 (96) |
| | Glyceryl palmito stearate (GPS) Myristic acid Poloxamer-407 (PL-407) | GPS (590 mg) MA (10 mg) PL-407 (480 mg) | 162±7.07 | 0.444±0.109 | - | 94±0.02 | |

| | | | | | | |
|---|---|---|---|---|---|---|
| Glyceryl palmito stearate (GPS) Myristic acid Poloxamer-407 (PL-407) | GPS (620 mg) MA (60 mg) PL-407 (550 mg) | 155±5.34 | 0.502±0.151 | - | 96.7±0.4 | |
| Glyceryl palmito stearate (GPS) Myristic acid Poloxamer-407 (PL-407) | GPS (600 mg) MA (10 mg) PL-407 (440 mg) | 166±1.83 | 0.177±0.221 | - | 86.6±0.06 | |
| Glyceryl palmito stearate (GPS) Myristic acid Poloxamer-407 (PL-407) | GPS (670 mg) MA (60 mg) PL-407 (530 mg) | 143±3.89 | 0.253±0.105 | - | 89.5±0.3 | |
| Glyceryl palmito stearate (GPS) Myristic acid Poloxamer-407 (PL-407) | GPS (730 mg) MA (90 mg) PL-407 (710 mg) | 129±1.83 | 0.177±0.221 | - | 93.1±0.6 | |

| | | | | | | | |
|---|---|---|---|---|---|---|---|
| Emulsion evaporation | Compritol-888 ATO (C-888) Gelucire (GL) Phospholipid (0.3%) S-100 (1% w/w) Pluronic F-68 (0.1% w/w) | C-888: GL = 1:1 | 205.1 ±9.3 | 0.38± 0.12 | -9.13± 0.97 | 72.3± 3.1 | Jain et al., 2018 (192) |
| | Compritol-888 ATO (C-888) Gelucire (GL) Phospholipid (0.3%) S-100 (1% w/w) Pluronic F-68 (0.1% w/w) | C-888: GL = 1:2 | 162.1 ±7.1 | 0.286 ±0.08 | -11.03 ±1.71 | 77.11 ±2.7 | |
| | Compritol-888 ATO (C-888) Gelucire (GL) Phospholipid (0.3%) S-100 (1% w/w) Pluronic F-68 (0.1% w/w) | C-888: GL = 2:1 | 185.1 ±9.3 | 0.192 ±0.07 | -9.48± 1.29 | 79.6± 2.9 | |
| High pressure homogenization | Modified starch (30% w/w) Medium chain triacylglycerol | Lycopene (0.1 %) | 151.2 ±2.56 | 0.211 ±0.014 | -19.83 ±0.16 | - | Li et al., 2018 (197) |
| | | Lycopene (0.3 %) | 146.87±2.12 | 0.213 ±0.023 | -19.73 ±0.31 | - | |
| | | Lycopene (0.5 %) | 145.1 ±2.31 | 0.218 ±0.04 | -20.23 ±0.40 | - | |
| Nano-precipitation | Acetone (5mL) Oligomerized-Epigallocatechin Polyethyl glycol sorbitan monooleate | Lycopene (11 w/w) | 152 | 0.21 | 58.3±4.2 | 89 | Li et al., 2017 (20) |
| | | Lycopene (15 w/w) | 225 | 0.20 | 50.9±3.5 | 85 | |
| | | Lycopene (20 w/w) | 276 | 0.26 | 44.2±4.7 | 81 | |
| High pressure homogenization | Tween-80 (3%) Lecithin (0.6%) | GMS (5%) | 183.4 ±3.03 | 0.32±0.009 | - | - | Zardini et al., 2017 (198) |
| | | GDS (5%) | 121.79±4.09 | 0.27±0.011 | - | - | |
| | | GMS (4%) MCT oil (1%) | 133.39±2.71 | 0.28±0.005 | - | - | |
| | | GDS (4%) MCT oil (1%) | 135.53±3.37 | 0.28±0.007 | - | - | |
| Ultrasonication | Solid lipid precirol ATO-5 (9-15%) Liquid lipid vit-E (4%) Tween-80:Ploxamer-188 (1:2) (2.5-7.5%) | | 121.9±3.66 | 0.370±0.97 | -29.0±0.83 | 84.50 | Sing et al., 2017 (39) |
| High pressure homogenization | Sesame oil + Lactoferrin (2%) | Sesame oil (0.1%) | 203.63±2.39 | 0.19±0.029 | - | 77.39±4.56 | Zhao et al., 2020 (199) |
| | | Sesame oil (0.2%) | 210.9±0.85 | 0.21±0.006 | - | 88.71±1.18 | |
| | | Sesame oil (0.4%) | 206.67±5.66 | 0.20±0.022 | - | 71.30±1.62 | |
| | | Sesame oil (0.6 %) | 216.7 ±2.61 | 0.22± 0.003 | - | 77.22 ±3.11 | |
| | | Sesame oil (0.8 %) | 210.53±3.80 | 0.22± 0.014 | - | 61.03 ±0.35 | |
| | | Sesame oil (1.0 %) | 213.2 ±4.11 | 0.21± 0.010 | - | 62.05 ±0.45 | |
| Rotary evaporation film ultrasonication method | Lecithin Cholesterol Dichloromethane | - | 58-105 | | -37 to -32.5 | | Zhao et al., 2018 (98) |
| Thin film hydration method | Soybean hosphatidylcholine (150 mg) Cholesterol (30 mg) Chloroform (5 mL) | - | 160.4 ±23.8 | | - | 81.85 | Zhuet. al. 2020 (200) |

Footnotes: GDS = Glycerol Distearate, GMS = Glycerol Monostearate, MCT oil = caprylic/capric triglyceride

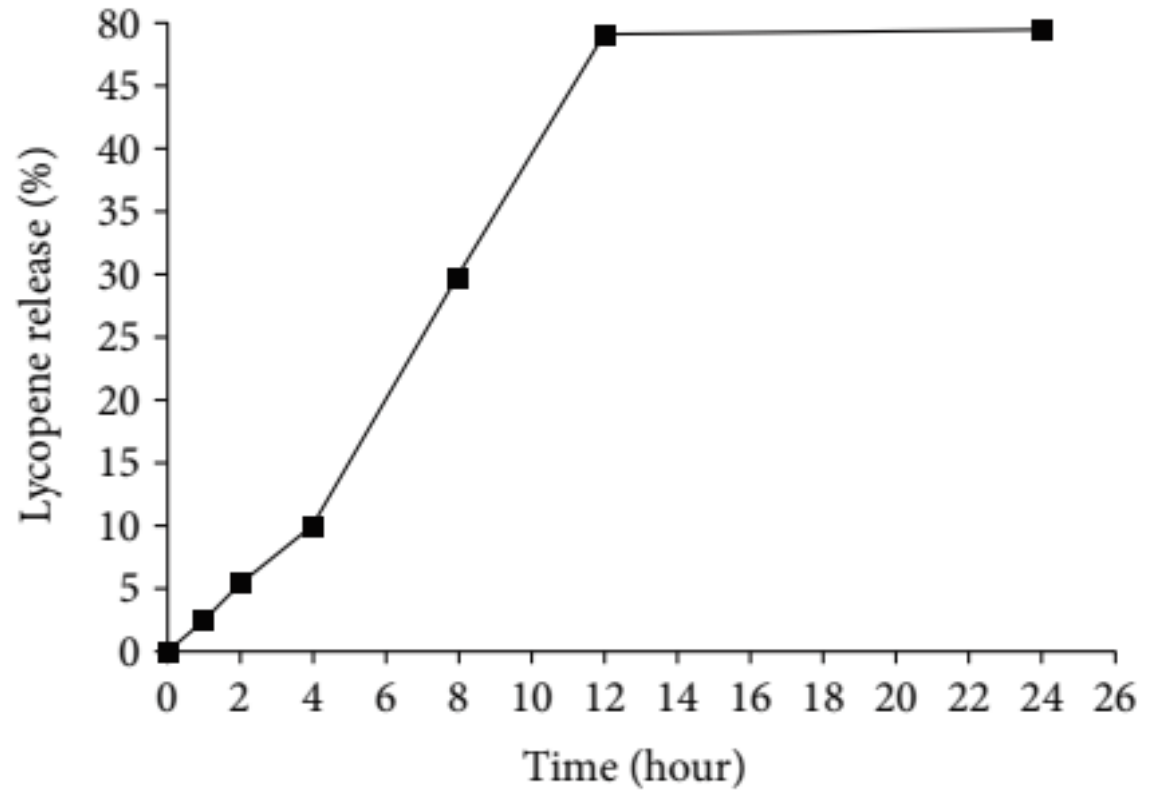

Figure-2 (a): Lycopene release from Lycopene nanoliposome (190)

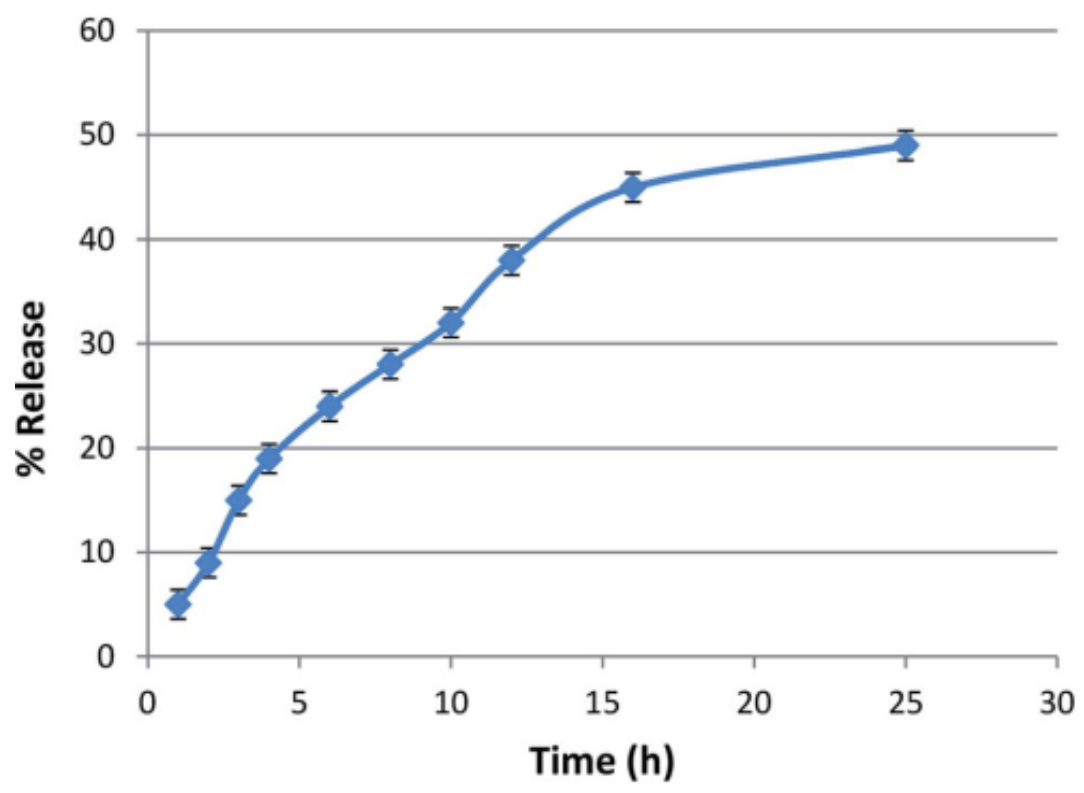

Figure- (b): Lycopene release from N-isopropylacrylamide-lycopene NPs (160)

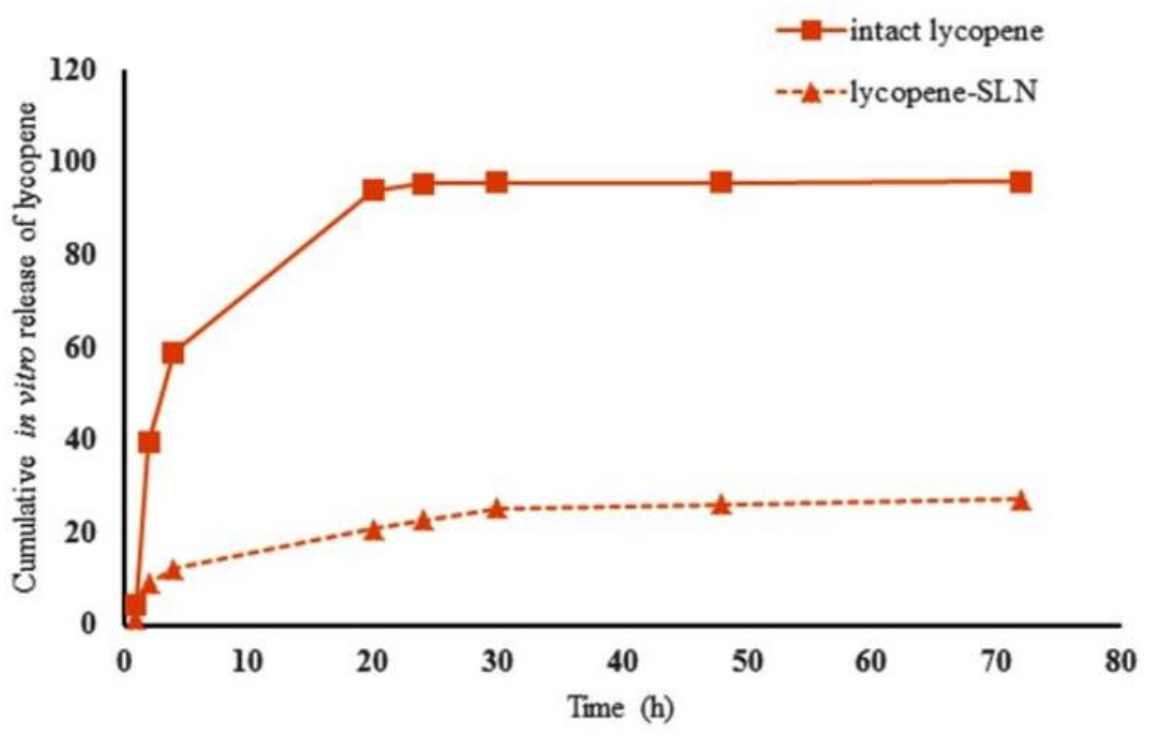

Figure-2 (c): Lycopene release from Lycopene solid lipid NPs (96)

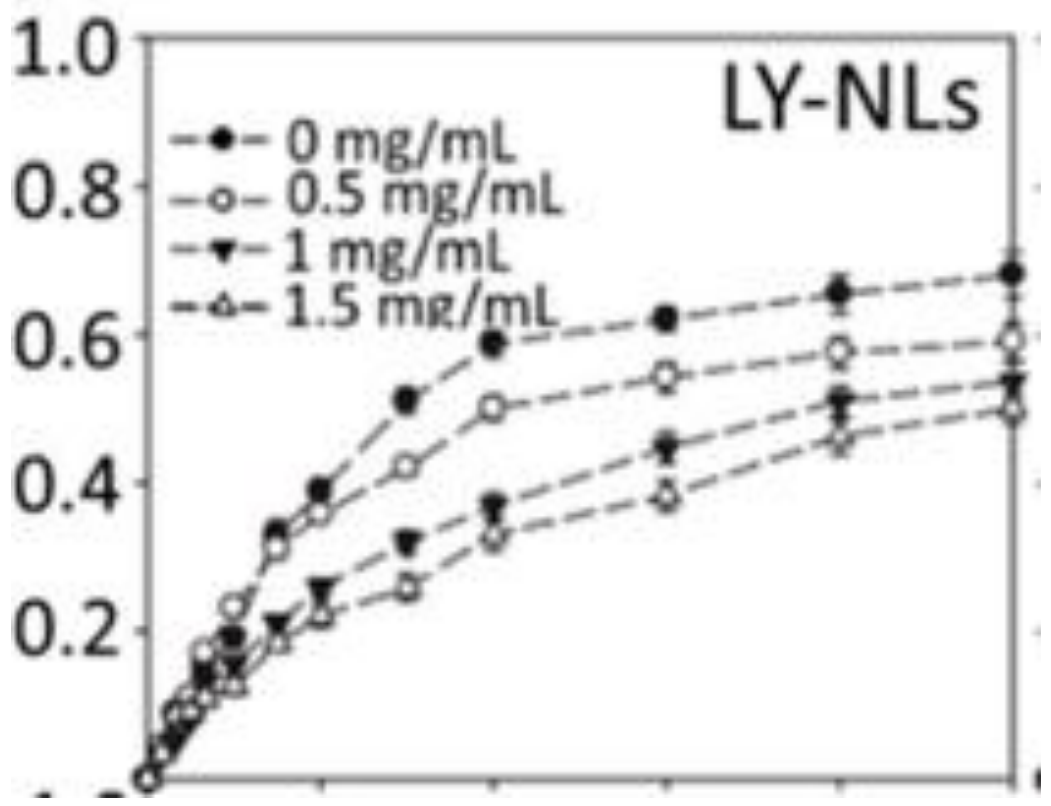

Figure-2 (d): Lycopene release from Polymeric (Chitosan)Lycopene NPs (201)

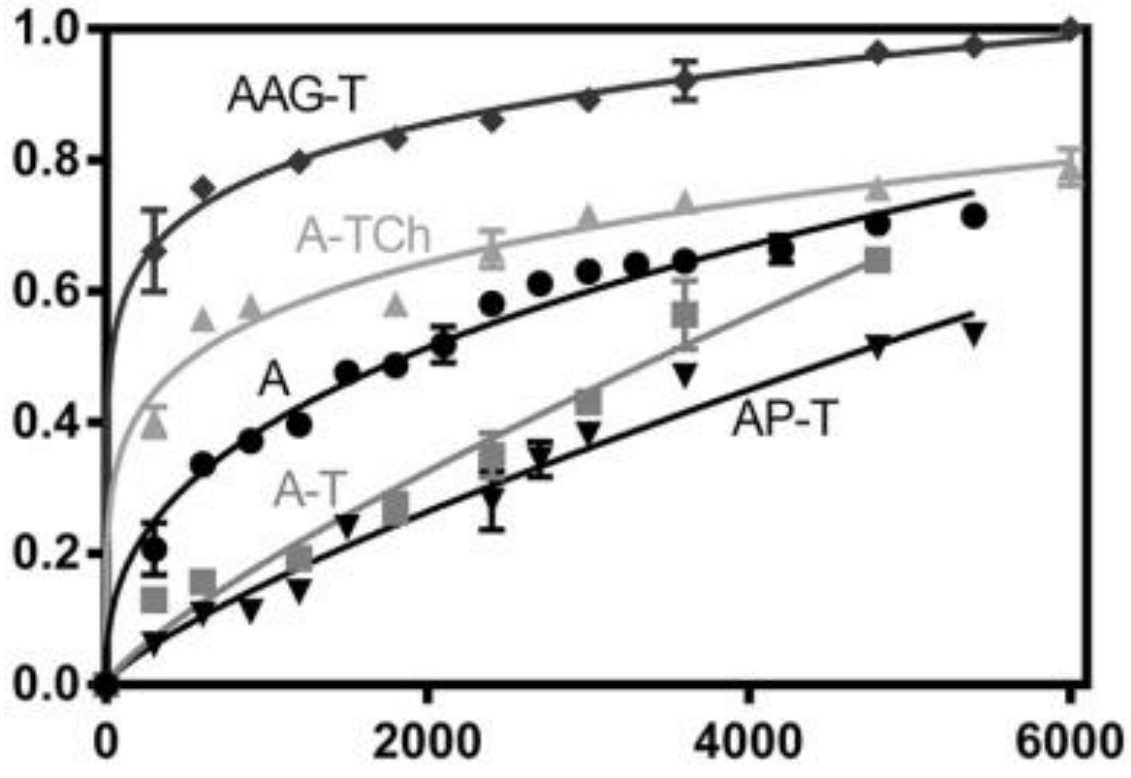

Figure-2 (e): Lycopene release from Polymeric (PLGA) Lycopene NPs (25) **(**Lycopene release (Lt/L∞) against time. Lines show the Peppas equation fitness. Standard deviation values are included. Elaboration of A = alginate, T = trehalose, Ch = chitosan, P = pectin, AAG = Arabic gum)

**Figure-2: Control release kinetics of lycopene nanoparticles**

Table-2: Interpretation of diffusion release mechanisms

| **Release exponent (n)** | **Drug transport mechanism** |
|---|---|
| n < 0.5 | Fickian diffusion |
| 0.45 < n < 0.89 | Non-Fickian transport |
| 0.89 | Cass II transport |
| 0.89 < n | Super case II transport |